\documentstyle[aps,prl,epsf,multicol]{revtex}

\title{Optimal Mutation Rates on Static Fitness Landscapes}

\author{Martin Nilsson}

\address
{Institute of Theoretical Physics, Chalmers University of Technology and G\"{o}teborg University, S-412 96 
G\"oteborg, Sweden {\tt martin@fy.chalmers.se}}

\date{\today}

\begin{document}
\maketitle

\begin{abstract}

We study the evolution of mutation rates for an asexual population living on a static fitness landscape,
consisting of multiple peaks forming an evolutionary staircase. The optimal mutation
rate is found by maximizing the diffusion towards higher fitness. Surprisingly the 
optimal genomic copying fidelity is given by $Q_{opt} = e^{ -\frac{1}{\ln \nu}}$ (where $\nu$ is the genome
length), independent of all other parameters in the model. Simulations confirm this
theoretical result. We also discuss the relation between the optimal mutation rate
on static and dynamic fitness landscapes.

\end{abstract}

\begin{multicols}{2}
\narrowtext

Evolution on the molecular level can be viewed as a diffusion process. The equations describing
the time dynamics of a population of gene sequences are a set of discrete diffusion 
equations with an exponential growth term. The diffusion stems from inaccurate copying of the genome
during replication. This enables the population to explore the sequence space, i.e., the 
space spanned by all possible gene sequences. Point mutations makes the Hamming distance a
natural metric on sequence space, which becomes topologically equivalent to a hyper-cube
of dimension $\nu$, where $\nu$ is the genome length. The high dimensionality makes 
analysis of the general diffusion process difficult. In this paper we focus on the evolution through a
specified path in the hypercube and disregard the dynamics of all other gene sequences. This 
gives a one dimensional sequence space. 
We are interested in the optimal mutation rate, which is defined as the mutation rate that maximizes
the diffusion speed.

The genome codes mainly for proteins which regulate
the chemical reactions within the cell. One of the processes that are under genomic control is the 
replication of the genome itself. When the genetic material is copied there are replicase 
enzymes involved. This is important since an unguided base pairing process is highly inaccurate. The
enzymes are determined by the genome and the mutation rate of the organism is therefore under 
evolutionary control. This implies that the mutation rates observed in living organisms have been selected for by 
Darwinian evolution. 

Naively one may think that since most mutations that affect the fitness are deleterious, organisms
should evolve as low mutation rates as possible. Measurments of mutation rates however show
that organisms have copying fidelities much below what could be expected from this 
assumption~\cite{DCCC98,Drake93}. 
They also show that the genomic mutation rate, i.e., the probability of one or more mutations to occur 
during one replication of the whole genome, is approximantly constant within similar groups
of organisms. This is surprising since the copying of the genetic material is a local process and
it is the mutation rate per base pair that are directly affected by the replicase enzymes.
Most attempts to find an evolutionary explanation for the observed mutation rates have been based on
populations evolving in a changing environment, see e.g.,~\cite{Kimura67,Leigh73,Ishii..89,Gillespie81,OM98,OF99,Ochoa..99a,opt-dyn}.
It is easy to understand that a non-zero mutation rate
is selected for on a dynamic fitness landscape, since perfect copying will unable adaption to new conditions. 
Recently a theoretical study has shown that the optimal genomic copying fidelity in a dynamic environment
is approximately independent of genome length~\cite{opt-dyn}. The theory also predicts mutation rates of the same order of magnitude
as observed for simple DNA based organisms. In this paper
we study a different model. The population lives in a static environment, but starts
far from the global fitness maximum. A non-zero mutation rate is selected for by 
maximizing the rate of evolution towards better fit genotypes. 

Consider an asexual haploid population of individuals, represented by genomes of length $\nu$. The fitness landscape consists of 
a number of peaks with superior fitness surrounded by a background. The evolution on this landscape
is driven by mutations enabling jumps from one fitness peak to a higher peak in 
the close neighborhood. We study a population of $N$ gene sequences starting at a low fitness peak which then mutate
onto successive fitness peaks of increasing height ($\sigma _1 < \sigma _2 < \cdots$). Furthermore we assume the copying fidelity 
per base, $q$, to be constant over the genome. The probability of a gene sequence to copy onto itself during 
one replication event, the genomic copying fidelity, is then given by $Q = q^{\nu}$. We also assume the probability of 
an individual on peak $\sigma_{i-1}$ to produce an offspring 
on peak $\sigma _i$ during a replication event to be $p_i (1-q)^{\alpha _i} q^{\nu - \alpha _i}$. This  
means that the number of bases where the sequences defining peak $\sigma _{i-1}$ and $\sigma _i$ differ is $\alpha _i$. The
factor $p_i$ is an arbitrary combinatorial factor, accounting for possible redundancies in sequence space, alphabet size, etc. All 
higher fitness peaks, $\sigma _k$ for $k > i$, are assumed to be further away so that mutations from peak $\sigma _{i-1}$ 
can be neglected. The evolution of the relative concentrations $x_n$ is described by
differential equation

\begin{eqnarray}
	\dot{x} _1 & = & W_{1, 1} \theta _N (x_1) + W_{1, 2} \theta _N (x_2) - f x_1 \nonumber \\
	\dot{x} _2 & = & W_{2, 1} \theta _N (x_1) + W_{2, 2} \theta _N (x_2) + W_{2, 3} \theta _N (x_3) - f x_2 \nonumber \\
	& \vdots & \nonumber \\
	\dot{x}_n & = &  W_{n, n-1} \theta _N (x_{n-1}) + W_{n, n} \theta _N (x_n) + \nonumber \\
	& & W_{n, n+1} \theta _N (x_{n+1}) - f x_n
\label{diffeq}
\end{eqnarray}
where the function $\theta _N$ is defined as $\theta _N (x_n) = x_n$ if $x> \frac{1}{N}$ and $0$ otherwise, and therefore accounts 
for the limited population size. The factor $f = \sum _i \left( q^{\nu}  + p_i (1-q)^{\alpha _i} \right) \sigma _i 
\theta _N (x_i)$ ensures $x_i$ to be normalized as relative concentrations. The matrix elements of $W$ are given by

\begin{eqnarray}
	W _{n, n} & = & q^{\nu} \sigma _n \nonumber \\
	W _{n-1, n} & = & p_{n} (1-q)^{\alpha _{n}} q^{\nu - \alpha _{n}} \sigma _{n-1} \nonumber \\
	W _{n, n+1} & = & p_{n+1} (1-q)^{\alpha _{n+1}} q^{\nu - \alpha _{n+1}} \sigma _{n+1} 
\end{eqnarray}
We start with a population that consists of individuals on the first peak $\sigma _1$, i.e., we define the
initial values as

\begin{eqnarray}
	x_i (0) & = & \left\{ \begin{array}{ll} 1 & i=1 \\ 0 & i \neq 1 \end{array} \right.
\end{eqnarray}

\begin{figure}[p]
\centering
\leavevmode
\epsfxsize = 0.8 \columnwidth
\epsfbox{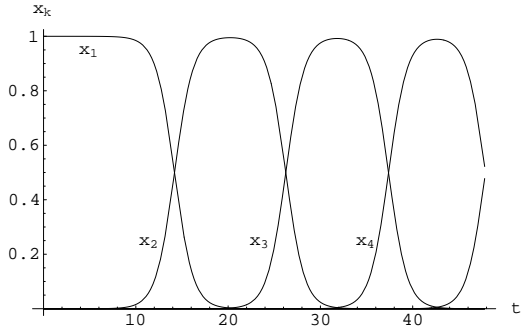}
\caption{\small The time dynamics of Eq.~\ref{diffeq} is simulated numerically. When the population diffuses off the 
initial peak $\sigma _1$ it starts evolving to peaks with higher and higher fitness. The parameters used
in this plot are $\nu = 100$, $\sigma _i = i$, $p = 0.01$, $Q = 0.99$ and $N = 10^6$.}
\label{dyn}
\end{figure}

The infinite population size limit of Eq.~\ref{diffeq} corresponds to a discrete normalized one-dimensional diffusion equation 
with an exponential growth term. However, this limit is not interesting for realistic systems since it does not allow 
propagating distributions
of concentrations localized in sequence space. If the fitness grows faster than linearly for example, 
the concentration on fitness peaks far from the starting point
grow large {\em before} the concentrations on peaks closer to the origin. This bizarre effect stems from the exponential growth
of very small (exponentially decaying with the distance from the origin) but non-zero concentrations
over all the fitness peaks shortly after the start.

In this model we implicitly assume the mutation rates to evolve much slower than 
the fitness, i.e. there are no significant changes in the mutation rate during the
evolution from one fitness peak to the next peak.

\begin{figure}[p]
\centering
\leavevmode
\epsfxsize = 0.8 \columnwidth
\epsfbox{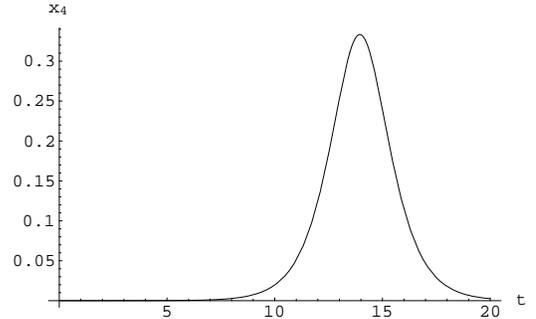}
\caption{\small  The picture shows $x_4$ given by Eq.~\ref{approx}.  The relevant parameter are the same as in
Fig.~\ref{dyn}; $\nu = 100$, $\sigma _i = i$, $p = 0.01$ and $Q = 0.99$. The maximum occurs at 
time $T \approx 14$ (defined as the time from the last peak's maximum). This is in agreement with the numerical
solutions shown in Fig.~\ref{dyn}.}
\label{pic-approx}
\end{figure}

The optimal copying fidelity $q_{opt}$ is defined by maximizing the diffusion speed
towards genotypes with superior fitness. 
Mathematically this corresponds to minimizing 
the time $T$ it takes for the concentration $x_n$ on peak $\sigma _n$ to reach its maximum, when the 
population starts at the proceeding peak $\sigma _{n-1}$. At the time when mutants from peak 
$\sigma _{n-1}$ have enabled the concentration $x_n$ to become large enough, i.e. $x_n > \frac{1}{N}$, exponential
growth will start with initial concentration proportional to $p_n (1-q)^{\alpha _n}$.  Since the population at this time 
is localized around peak $n$ the concentration $x_n$ is described approximately by

\begin{eqnarray}
	x_n ( t) & \sim & \frac{ \gamma e^{q ^{\nu} \sigma _n t}}{e^{q ^{\nu} \sigma _{n-1} t} +
		\gamma e^{q ^{\nu} \sigma _n t} + \gamma ^2 e^{q ^{\nu} \sigma _{n+1} t}}
\label{approx}
\end{eqnarray}
where $\gamma = p_n (1-q)^{\alpha _n}$. The denominator normalizes $x_n$ by summing the absolute growth
in the surrounding of peak $n$, see Fig.~\ref{pic-approx}. The time $T$ when
$x_n (t)$ has a maximum can be found by solving $\frac{d x_n (t)}{d t} = 0$, giving

\begin{eqnarray}
	T & = & - \frac{1}{\sigma _{n+1} - \sigma _{n-1}} \cdot \frac{\ln \left( \gamma ^2 \kappa \right)}{q^{\nu}}
\label{time}
\end{eqnarray}
where $\kappa = \frac{\sigma _{n+1} - \sigma _n}{\sigma _n - \sigma _{n-1}}$. The diffusion speed is defined as
$v = \frac{1}{T}$. By making the approximation $\kappa \approx 1$, we can write

\begin{eqnarray}
	V & = & - \frac{\sigma _{n+1} - \sigma _{n-1}}{2} \cdot \frac{q^{\nu}}{\ln \left( \gamma \right)}
\label{diff_speed}
\end{eqnarray}
%It is interesting to note that the diffusion speed, i.e. the rate of evolution, is proportional to
%the local selection pressure, this was first noted in a model by Kimura.

\begin{figure}[p]
\centering
\leavevmode
\epsfxsize = 0.8 \columnwidth
\epsfbox{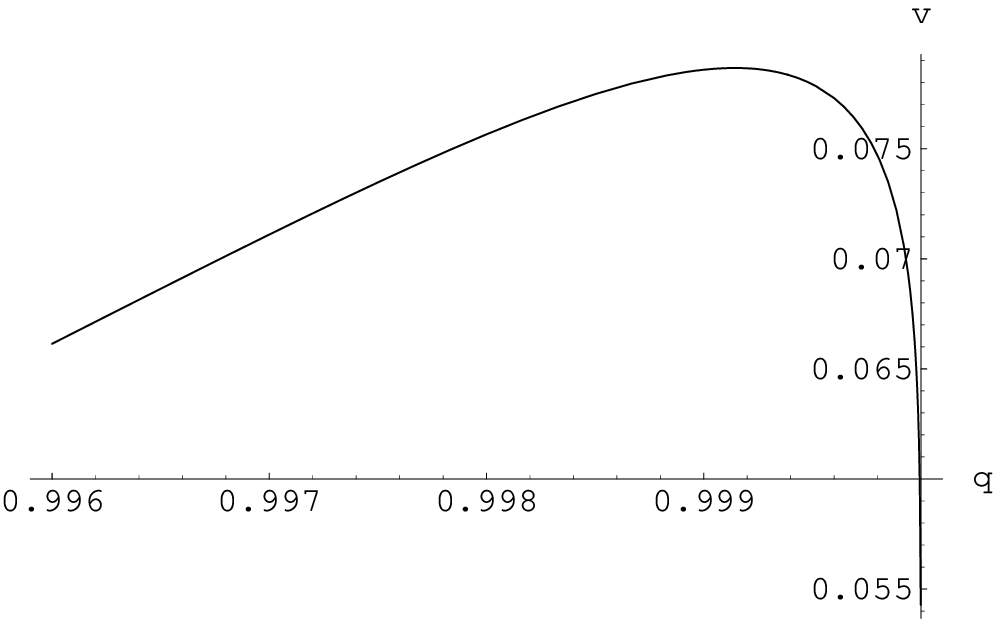}
\caption{\small The figure shows $V (q)$ given by Eq.~\ref{diff_speed}. The maximum 
gives the optimal copying fidelity $q_{opt}$. Parameters used in the figure are $\nu = 100$, 
$\sigma _i = i$ and $p = 0.01$. The shape of the cure is not
sensitive to the parameter values, as long as $\nu \gg 1$.}
\label{fig_speed}
\end{figure}

The optimal copying fidelity $q_{opt}$ is defined to maximize the diffusion speed, and can therefore
be derived by finding the maximum of $V (q)$ in Eq.~\ref{diff_speed}
(see Fig.~\ref{fig_speed}). Setting the derivative to zero, $\frac{d V}{d q} =0$, and noting that $q \approx 1$ 
gives the equation:

\begin{eqnarray}
	1 + \frac{1}{\nu (1-q) \left( \ln ( \bar{p}) + \ln  (1-q) \right)} & = & 0
\label{eq}
\end{eqnarray}

\begin{figure}[p]
\centering
\leavevmode
\epsfxsize = 0.8 \columnwidth
\epsfbox{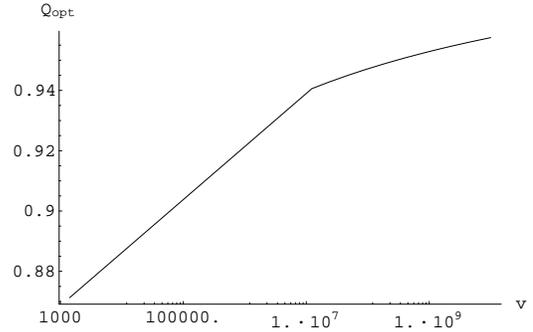}
\caption{\small The figure demonstrates how weakly $Q_{opt}$ scales with genome length. }
\label{insen}
\end{figure}

We are interested in the limit where the genome length is large. 
In this limit the first term in the denominator (involving $\bar{p}$) can be neglected. Eq.~\ref{eq} then
reduces to
\begin{eqnarray}
	\nu (1-q) \ln (1-q) & = & -1
\label{eq_mu}
\end{eqnarray}
There is no closed analytic expressions for the solution to this equation, but a converging iterative
expression can be found for the optimal copying fidelities
\begin{eqnarray}
	q _{opt} & = & 1 - \frac{1}{\nu \ln \left( \nu \ln \left( \nu \ln \left( \cdots \right) \right) \right)} \nonumber \\
	Q _{opt} & \approx  & e ^{- \frac{1}{\ln \nu}}
\label{main_result}
\end{eqnarray}
It is surprising that the optimal genomic copying fidelity depends so weakly on the genome length,
and even more surprising that it is independent of all other parameters in the model. This independence is both interesting and
important, especially since we start by assuming a specific path for evolution. As it turns out the 
optimal mutation rate does not depend on the particular path chosen.  The insensitivity of $Q_{opt}$ when the genome length 
varies can be seen by considering biologically plausible genome lengths, see Fig.~\ref{insen}. 
Note that the genomic copying fidelity increases with genome length.

\begin{figure}[p]
\centering
\leavevmode
\epsfxsize = 0.8 \columnwidth
\epsfbox{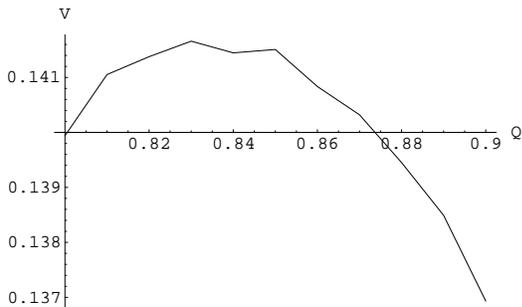}
\caption{\small The figure shows the region where $V (q)$ has a maximum, calculated by numerical simulations 
of Eq.~\ref{diffeq}. Parameter settings in the simulations were  $p = 0.01$, $\alpha = 1$, $N = 10^8$ and
$\nu = 1000$. The minimum occurs approximately at the point predicted by Eq.~\ref{main_result}, 
i.e., $Q_{opt} = 0.86$.}
\label{diff_time}
\end{figure}

In simulations of a population consisting of $2000$ individuals with genome length $\nu = 70$ on a rugged 
fitness landscape (created by an elementary folding algorithm for calculating secondary structures
of gene sequences), Fontana and Schuster~\cite{fontana87} find
that the rate of evolution is maximal approximately at $\mu = 0.003$. This is 
in close agreement with the mutation rate as predicted by Eq.~\ref{main_result} for genome length $70$, $\mu _{opt} = 0.0025$.

The optimal copying fidelity given in Eq.~\ref{main_result} can also be  derived using a 
more intuitive argument. The argument also shows
more clearly how evolved mutation rates on static fitness landscapes relate to
evolved mutation rates in dynamic environments. The rate of growth between two peaks, with fitness
difference $\Delta \sigma$, is given by
$e^{Q \Delta \sigma t}$. The diffusion from an occupied peak to the next is proportional to
$(1-q)^{\alpha}$, where $\alpha$ measures the distance in sequence space between the peaks. The 
time, $T$, it takes for a population to evolve from one peak to an other will therefore be given by
the solution of the equation $(1-q)^{\alpha} e^{Q \Delta \sigma t} = 1$, i.e. 
$T \sim - \frac{\alpha \ln (1-q)}{q^{\nu} \Delta \sigma}$. 
Organisms, free to change their mutation rates, evolve a copying fidelity $q_{opt}$ that minimizes $T(q)$. 
Deriving an expression for the equation $\frac{d T}{d q} = 0$, using 
$q^{\nu} \approx 1$, gives $\frac{1}{1-q} + \nu \ln (1-q) = 0$, which is equivalent
to Eq.~\ref{eq_mu} and is solved by $Q_{opt} \approx  e ^{- \frac{1}{\ln \nu}}$. 

In a recent paper~\cite{opt-dyn}, the evolution of mutation rates on a dynamic fitness
landscape was studied. The fitness landscape consists of a single peak moving around in
sequence space, shifting position on average once every $\tau$ generations. The relative 
selective advantage for a sequence on the fitness peak is $\sigma$. A shift of the peak consist of
$\alpha$ changes of bases in the sequence defining the fitness peak. Since an individual in the 
population needs to produce offspring that are able to follow the shifts of the fitness peak, a 
non-zero mutation rate is selected for. It turns out that finding the optimal copying fidelity
is equivalent to minimizing $(1-q)^{\alpha} e^{Q \sigma \tau}$ with respect to $q$. This is the same 
expression as for the growth rate between fitness peaks on a static landscape. However, in 
the dynamic case the growth over
a cycle, consisting of a shift and a static period, is be optimized rather than the time to
evolve from one peak to the next. More generally, if the evolution of mutation rate is driven by
a dynamic environment it will be selected to optimize the growth on the changing fitness landscape,
whereas on a static landscape the mutation rate maximizing the rate of evolution towards higher
fit genotypes will be selected for. Maximizing the growth on a dynamic landscape gives
$Q_{dyn} = e^{- \frac{\alpha}{\sigma \tau}}$.

There are some fundamental differences between the two models presented above. In the model based on dynamic fitness landscapes
the population dynamics is driven by external changes of the environment. The organisms have to passively wait for the environment
to change and then adapt to the new fitness landscape. In the model based on rugged fitness landscapes the situation is different.
There always exist a higher fitness peak in the close neighborhood so the population has to minimize the time for diffusing
to and growing large on the higher peak. Hence the population should actively search the surroundings in sequence space. 
The main difference between the models is therefore the preexistent of higher fitness peaks close in sequence space,
which results in very different optimal mutation rates.

The genomic copying fidelity in both the static and dynamic case is approximately independent
of genome length, a phenomenon that is also observed in nature. To be more precise, experiments show 
that the genomic copying fidelity is approximately constant within groups of similar
organisms, e.g., simple DNA-based organisms have $Q \approx 0.996$ whereas RNA based retroviruses
have $Q \approx 0.9$~\cite{DCCC98}.  Simple DNA based organisms for example have
much too low mutation rates to be explained by evolution on the static landscapes studied
in this paper. Retroviruses on the other hand show mutation rates that are in agreement with
the predictions made in this paper. However, they may also be explained by mutation rates evolved as a response to
a changing environment as discussed above. It is therefore unclear whether the major force behind
the evolution of mutation rates for retro virus is maximizing the evolution rate towards higher fitness or
maximizing the growth in a changing environment (caused be the immune system).

In conclusion, we show that the optimal genomic copying fidelity, i.e., that which optimizes the rate of evolution, on a rugged
fitness landscape can be written as $Q _{opt} = e^{-\frac{1}{\ln (\nu)}}$, where $\nu$ is the genome length. The optimal
genomic copying fidelity on rugged fitness landscapes is predicted to be around $0.9$ for realistic
genome lengths ($\nu \in  [10^3, 10^{10} ]$). Of the mutation rates observed in nature, retroviruses (including HIV)
confirm this prediction. The model presented here therefore presents a possible explanation for the observed
mutation rates for retro viruses.

The author would like to thank Jennie Jacobi and ``Mullb\"aret'' for providing a nice and
stimulating environment while working on the ideas behind this paper. Thanks are also due to
Mats Nordahl and Johan Ivarsson for valuable comments on the manuscript.

\end{multicols}

\end{document}